\pdfoutput=1
\documentclass[a4paper,11pt]{article}
\usepackage[T1]{fontenc}
\usepackage[utf8]{inputenc}
\usepackage{lmodern,amsmath,amssymb,graphicx}
\usepackage[margin=25mm]{geometry}
\usepackage{hyperref}
\usepackage{float,booktabs,tabularx,microtype}
\title{Fast Monte Carlo calculation of multiple scattering in resonant neutron capture}
\author{R. M. Djilkibaev\\[4pt]\small Institute for Nuclear Research of the Russian Academy of Sciences\\\small 7a, 60-letiya Oktyabrya Prospekt, Moscow 117312, Russia\\\small\texttt{rmd@inr.ru}}
\date{}
\begin{document}
\maketitle
\begin{abstract}
A fast Monte Carlo code has been developed to calculate multiple neutron scattering in a sample. Good agreement with an independent Monte Carlo calculation using SAMMY is obtained for neutron capture in $^{197}$Au over the energy range 50--300 eV. For capture following two or more scattering events, a difference is found between the direct Monte Carlo calculation and the analytical approximation in SAMMY. The computational speed allows the scattering correction to be recalculated whenever trial resonance parameters are changed during parameter estimation. The computational core is implemented in C++, while data preparation, calculation control and analysis are performed in Python. Model studies examine how Monte Carlo statistics affect the accuracy with which neutron and radiative resonance widths are recovered.
\end{abstract}
\section{Introduction}

In neutron capture measurements, some neutrons are captured after undergoing one or more scattering events within the sample. Scattering changes the neutron energy and direction and, consequently, the probability of subsequent capture. These events contribute to the capture yield and modify its energy dependence near resonances. Determining resonance parameters therefore requires accounting for both capture without prior scattering and capture following scattering.

When resonance parameters are determined from a capture yield curve, their trial values are varied. This changes the interaction cross sections and, consequently, the probability of neutron capture following scattering. The contribution of these events must therefore be recalculated during parameter estimation. Monte Carlo methods account for successive scattering events and changes in neutron energy, but repeated calculations may require substantial computing time. Practical applications require a fast calculation with sufficiently low statistical noise to preserve the accuracy of the estimated resonance parameters.

Mancinelli \cite{ref1} describes a weighted Monte Carlo method for calculating multiple neutron scattering in a sample. Rather than simulating a large number of neutrons that leave the sample without interacting, the method samples a collision within the sample and accounts for its probability through a statistical weight. That work concerns neutron diffraction and treats scattering without a change in energy. Resonant capture calculations must additionally account for neutron energy loss during scattering and the resulting change in the probability of subsequent capture.

The resonance analysis code SAMMY \cite{ref2} accounts for neutron scattering within the sample. In addition to analytical approximations, it provides a Monte Carlo calculation that simulates successive neutron interactions in the target. Comparison with this calculation allows a new code to be checked under the same physical conditions and with the same input resonance parameters. In this work, SAMMY MC provides an independent calculation for validating our code, hereafter referred to as MC-fast.

The importance of multiple scattering in resonance analysis is illustrated by neutron capture measurements on $^{50}$Cr and $^{53}$Cr at n\_TOF \cite{ref3}. The authors used samples of different thicknesses and determined the parameters of strong resonances from thin samples to reduce the scattering contribution. In thicker samples, capture following scattering accounted for more than half of the total capture yield in some regions. SAMMY calculations using parameters obtained from the thin samples did not reproduce the measured capture yield in the thick samples. This example demonstrates the importance of accurate multiple scattering (MS) calculations when determining resonance parameters.

Nobre et al. \cite{ref4} reevaluated neutron data for chromium isotopes, including an updated R-matrix analysis of strong resonances in $^{50}$Cr and $^{53}$Cr. The resulting data were tested against integral experiments sensitive to neutron interactions with chromium.

The result of the present work is a fast implementation of the MS calculation, validated against SAMMY MC, that allows the correction to be recalculated whenever trial widths are changed during resonance parameter recovery.

\section{Monte Carlo method}

\subsection{Neutron capture probability in the target}

\begin{figure}[H]
\centering
\includegraphics[width=.92\linewidth]{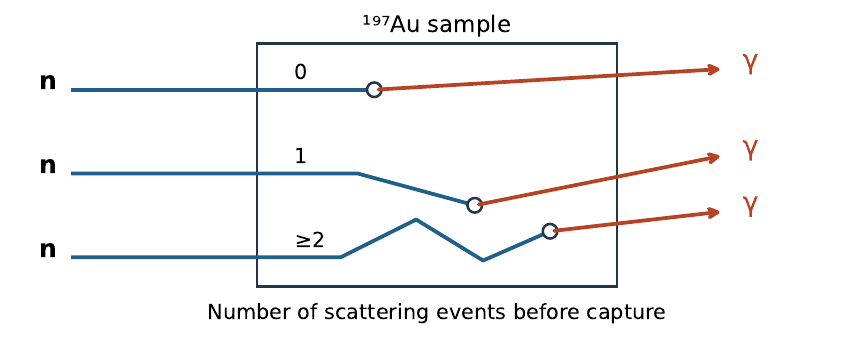}
\caption{A neutron can be captured directly or after one or more scattering events. The contributions to the capture yield are separated according to the number of scattering events preceding capture.}
\label{figure:1}
\end{figure}

Y(E) is the probability that a neutron with incident energy E is captured in the target, referred to here as the capture yield. It is separated into three contributions: capture without prior scattering, $Y_0$; capture after one scattering event, $Y_1$; and capture after two or more scattering events, $Y_{\geq2}$:
\begin{equation}
Y(E)=Y_0(E)+Y_1(E)+Y_{\geq2}(E).
\end{equation}
The sum $Y_1$ + $Y_{\geq2}$ is referred to below as the multiple scattering (MS) correction. These processes are illustrated in figure~\ref{figure:1}.

The calculation considers a gold target modelled as an infinite planar slab, with neutrons incident normally on its surface. Two areal densities are considered: $n_0$ = $7.3\times10^{-4}$ atoms/barn for the n\_TOF example \cite{ref5} and $n_0$ $\approx$ $6.3\times10^{-3}$ atoms/barn for INES \cite{ref6}. The areal density specifies the number of atoms per unit target area and, together with the cross sections, determines the neutron interaction probability. These values allow the calculation to be tested for substantially different contributions from repeated scattering. The resolution function and other facility-specific effects are not included in these tests.

The probability of capture without prior scattering is calculated directly from the cross sections:
\begin{equation}
Y_0(E)=[1-\exp(-n_0\sigma_{\mathrm{tot}}(E))]\frac{\sigma_\gamma(E)}{\sigma_{\mathrm{tot}}(E)}.
\end{equation}
Here $\sigma_{\mathrm{tot}}$ is the total interaction cross section and $\sigma_\gamma$ is the capture cross section. The expression in square brackets gives the probability of at least one interaction during passage through the target, while the cross-section ratio gives the probability that the first interaction is capture. Thus, $Y_0$ is calculated analytically, and Monte Carlo simulation is used to determine capture following scattering.

Calculating capture following scattering requires tracking the neutron within the target. Each scattering event changes its energy and direction. The new energy determines the interaction cross sections, while the neutron position and direction determine the distance to the target surface. These quantities govern the probability of the next interaction and subsequent capture. Monte Carlo simulation accounts for these changes by generating individual neutron histories.

MC-fast uses a weighted Monte Carlo method to make more efficient use of the simulated histories. Each history is assigned a statistical weight that accounts for the probability of the simulated sequence of events. At the sampled interaction point, the capture contribution is scored, and the neutron history continues after scattering with a reduced weight. A single history can therefore contribute to capture after different numbers of scattering events. Averaging these contributions over all histories gives the MS correction.

\paragraph{Main algorithm equations.}

Let D be the slab thickness and $n_0$ its areal density. The neutron enters normally with initial weight w = 1. The attenuation coefficient and mean free path at the current energy E are

\begin{equation}
\mu(E)=\frac{n_0}{D}\sigma_{\mathrm{tot}}(E),\qquad\lambda(E)=\frac{1}{\mu(E)}.
\end{equation}

If L is the distance to the surface along the current direction, the probability of an interaction before leaving the slab is
\begin{equation}
P=1-\exp[-\mu(E)L],\qquad\mu(E)=\frac{n_0}{D}\sigma_{\mathrm{tot}}(E).
\end{equation}
The distance $\ell$ to the collision is sampled conditional on the collision occurring within the slab:
\begin{equation}
\ell=-\frac{\ln(1-\xi P)}{\mu(E)}.
\end{equation}
where $\xi$ is uniformly distributed in (0, 1). To account for the collision probability, the current history weight w is multiplied by P. Denote the resulting weight by $w_c$ = w $\times$ P. At the collision point, the capture contribution and the weight for continuing the history after scattering are

\begin{equation}
\Delta Y=w_c\frac{\sigma_\gamma(E)}{\sigma_{\mathrm{tot}}(E)},\qquad w'=w_c\left[1-\frac{\sigma_\gamma(E)}{\sigma_{\mathrm{tot}}(E)}\right].
\end{equation}

The contribution $\Delta Y$ is assigned according to the number of scattering events preceding that collision. For the first collision, the mean capture contribution is already included in the analytical $Y_0$; only contributions following at least one scattering event are included in the MS correction. The history continues with the new weight $w'$, energy and direction. Contributions from a single history to different scattering orders are correlated, and this covariance is included when estimating the uncertainty of their sum.

The calculation considers elastic neutron scattering from gold nuclei. The scattering direction is sampled isotropically in the centre-of-mass frame. The neutron energy and direction after the collision are determined by conservation of energy and momentum for an initially stationary nucleus. A temperature of 295 K is included in the calculation of Doppler-broadened cross sections. Thus, nuclear thermal motion is accounted for in the cross sections but is not sampled separately at each collision.

Let $\varepsilon$ = cos $\theta_{\mathrm{cm}}$, where $\theta_{\mathrm{cm}}$ is the scattering angle in the centre-of-mass frame. The energy $E'$ and the angle $\theta$lab relative to the neutron direction before the collision are given by

\begin{equation}
E'=E\frac{A^2+2A\varepsilon+1}{(A+1)^2},\qquad\cos\theta_{\mathrm{lab}}=\frac{1+A\varepsilon}{\sqrt{A^2+2A\varepsilon+1}}.
\end{equation}

where A = 195.27400019 is the ratio of the gold atom mass to the neutron mass. The variable $\varepsilon$ is uniformly distributed in [-1, 1], and the azimuthal angle $\varphi$ in [0, 2$\pi$]. After the energy changes, the cross sections are evaluated at the new energy $E'$. Capture and total interaction cross sections between adjacent tabulated nodes are obtained by linear interpolation in energy.

All contributions to the capture yield are assigned to the neutron energy at entry into the target. The energy after scattering is used to calculate the probability of subsequent interactions but does not change the energy point to which the history contributes. A neutron entering above a resonance and losing energy through scattering can therefore be captured near the resonance and increase the capture yield at its incident energy. This explains the contribution from capture following scattering on the higher-energy side of the resonance peak.

When trial resonance parameters change, the capture and scattering cross sections are recalculated, followed by the corresponding capture yield. The MS correction therefore changes with the parameters and accounts for their effect on successive neutron interactions within the target. Even when parameters are determined within a narrow energy window, the cross sections are calculated using the full resonance set, because neighbouring resonances may affect interactions both before and after the neutron energy changes. In the tests considered here, only the widths of the selected resonance or resonance pair are varied; all other parameters remain fixed.

The number of simulated histories determines both the statistical precision of the MS correction and its computing time. As the number of independent histories N increases, the statistical uncertainty decreases approximately as $1/\sqrt{N}$, while the computing time under otherwise unchanged conditions increases approximately in proportion to N. Halving the uncertainty therefore requires roughly four times as many histories. The required statistics should be chosen according to the accuracy of resonance parameter recovery: noise in the MS correction affects the recovered widths through the sensitivity of the capture yield to those parameters.

The main calculations include capture following no more than five scattering events. This limit reduces computing time but requires checking the contribution from further scattering. The calculation is therefore repeated with a limit of ten scattering events, and the change in capture yield is compared with statistical noise. In the 50--300 eV comparison, histories with statistical weights below $10^{-7}$ are terminated. In the scattering-limit tests and parameter recovery calculations, such histories are treated using Russian roulette: some histories are randomly terminated, while the weights of the surviving histories are increased to preserve the mean capture contribution.

In MC-fast, the weighted neutron transport algorithm is implemented in a C++ computational core. Cross-section preparation, calculation control and resonance parameter recovery are performed in Python. Independent histories are distributed among worker processes, allowing several processor cores to be used. The speed of this implementation allows the MS correction to be recalculated whenever trial widths are changed during parameter recovery.

\section{Comparison with an independent SAMMY calculation}

MC-fast is validated by comparison with the independent Monte Carlo calculation in SAMMY. The local tests use a common cross-section table, allowing neutron transport and interactions within the target to be tested separately. In the broad 50--300 eV comparison, each code calculates its own cross sections from the same resonance parameter set, so the consistency of the cross-section calculations is also tested.

\subsection{Comparison conditions}

Both codes use the same set of $^{197}$Au parameters. The neutron and radiative widths correspond to ENDF/B-VIII.1 \cite{ref7}, whereas the resonance energies were previously adjusted in an analysis of n\_TOF data. These energies remain fixed in the present model tests. SAMMY version 8.2 is used for the comparison. MC-fast calculates cross sections in the Reich--Moore formalism with Doppler broadening at 295 K.

In the interval 240--241 eV, the MC contribution from capture following two or more scattering events is 48\%, 54\% and 72\% of the corresponding analytical SAMMY contribution for target thicknesses of 0.00610, 0.01220 and 0.02440 cm, respectively. Their areal densities are $3.65\times10^{-4}$, $7.30\times10^{-4}$ and $1.46\times10^{-3}$ atoms/barn, respectively. For the nominal thickness of 0.01220 cm, the MC value is approximately 46\% below the SAMMY result. The compared contributions are averaged over the energy interval.

In the analytical SAMMY calculation, capture following two or more scattering events is evaluated assuming a uniform neutron distribution throughout the target volume and isotropic directions after the second scattering event \cite{ref2}. Direct Monte Carlo simulation retains the actual neutron positions and directions, which determine the distance to the surface and the probability of subsequent capture. To investigate the discrepancy, a separate MC calculation evaluates the neutron escape probability after the second scattering event using the SAMMY approximation. The sampling of the first two scattering events and the cross sections remain unchanged. This diagnostic calculation uses 1.5 million histories for each target thickness. Its contribution differs from the analytical SAMMY result by 1.2\%, 0.9\% and 0.4\% for half the nominal thickness, the nominal thickness and twice the nominal thickness, respectively, averaged over 240--241 eV. These values describe the observed agreement; statistical uncertainties of the diagnostic calculation are not included in this assessment. The result points to the approximate treatment of neutron escape after the second scattering event as the main source of the observed discrepancy.

These percentages refer to capture following two or more scattering events, which contributes little to the total signal in this interval. Replacing only this SAMMY contribution with the MC result while retaining the other contributions would reduce the total capture yield by approximately 0.0042\%, 0.0111\% and 0.0186\% for half the nominal thickness, the nominal thickness and twice the nominal thickness, respectively. A large relative difference in this contribution therefore does not by itself imply a similarly large change in the total capture yield or the recovered resonance widths.

The diagnostic test explains the difference from the analytical SAMMY approximation under the conditions considered. Further validation of MC-fast uses SAMMY MC, which simulates neutron transport directly. Agreement between the two calculations tests the algorithm implementation under common physical assumptions; comparison with experimental data is outside the scope of this test.

The comparison is performed separately for capture without prior scattering, $Y_0$; after one scattering event, $Y_1$; and after two or more scattering events, $Y_{\geq2}$. This tests both the total capture yield and its distribution among different scattering orders. Since the dominant $Y_0$ contribution may conceal differences in the smaller MS correction, the sum $Y_{\geq1}$ = $Y_1$ + $Y_{\geq2}$ is also examined separately. The difference between this sum in the two calculations, divided by the total capture yield, shows how the discrepancy in MS affects the predicted yield.

The difference between two Monte Carlo calculations contains statistical noise from both codes. A small discrepancy at individual energy points therefore does not by itself imply a difference between the algorithms. The comparison is repeated with increased numbers of histories and independent random number sequences. A reduction in the scatter approximately as $1/\sqrt{N}$ and the absence of a persistent structure in the difference indicate statistical noise.

\subsection{Capture yield contributions near 240 eV}

The resonance near 240 eV is selected for a detailed comparison. It is sufficiently isolated to facilitate comparison of the shapes of $Y_0$, $Y_1$ and $Y_{\geq2}$. In addition, the model tests show that recovery of its widths is more sensitive to statistical noise in the MS correction than recovery of the selected 163--165 eV resonance pair. The full resonance set is used to calculate cross sections in both cases.

MC-fast uses 10 million histories with incident energies of 230--250 eV and averages the capture yield over energy bins of width 0.05 eV. SAMMY MC uses 1 million histories at each specified energy; its results are interpolated to the bin centres. The comparison therefore includes an additional difference between the mean yield within a bin and the yield at its centre, particularly near the rapidly varying resonance peak. In the subsequent tests, both codes calculate the yield at identical incident energies, removing this difference.

The comparison near the 240 eV resonance tests capture after one scattering event, $Y_1$, and after two or more scattering events, $Y_{\geq2}$, separately. Over 235.025--244.925 eV, the integrals of these contributions calculated by MC-fast and SAMMY MC differ by 0.44\% and 0.36\%, respectively (table~\ref{table:1}). Their effect on the total capture yield is smaller because capture without prior scattering, $Y_0$, is dominant here.

\begin{table}[H]
\centering
\caption{Integrals of the capture yield contributions, $\int Y_i(E)\,dE$ [eV], over the common window 235.025--244.925 eV. The quantities $Y_i$(E) are dimensionless.}
\label{table:1}
\small
\begin{tabularx}{\linewidth}{lXXX}
\toprule
Component & MC-fast, $\int Y_i(E)\,dE$ [eV] & SAMMY MC, $\int Y_i(E)\,dE$ [eV] & Ratio MC-fast / SAMMY MC \\
\midrule
$Y_0$ & 0.2887850 & 0.2888873 & 0.999646 \\
$Y_1$ & 0.00873982 & 0.00870180 & 1.004370 \\
$Y_{\geq2}$ & 0.000176046 & 0.000175408 & 1.003636 \\
\bottomrule
\end{tabularx}
\end{table}

\begin{figure}[H]
\centering
\includegraphics[width=.88\linewidth]{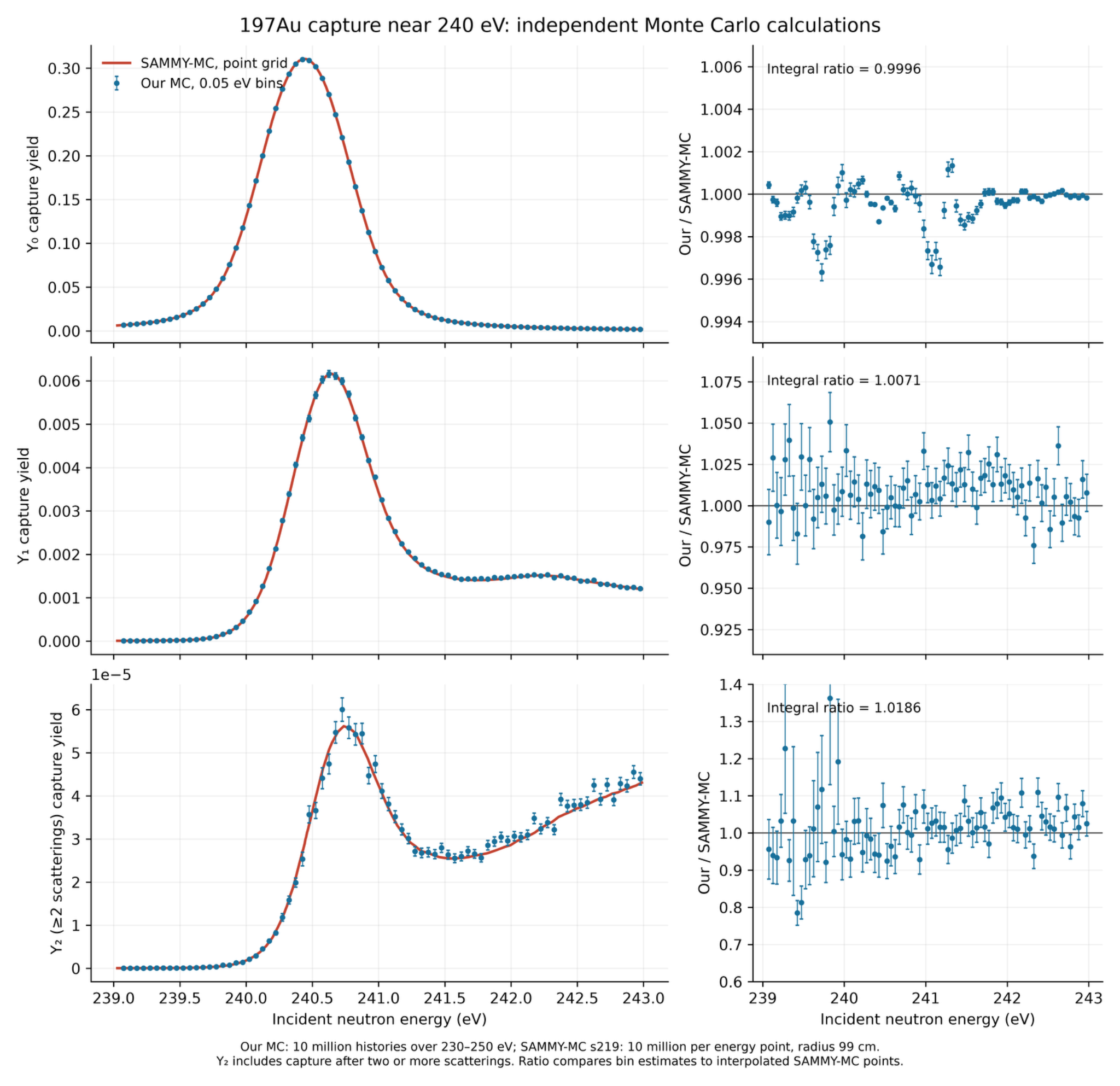}
\caption{Capture yield curves and their ratios near the peak of the 240 eV resonance, over 239--243 eV. The SAMMY MC calculation uses 10 million histories per incident energy.}
\label{figure:2}
\end{figure}

Figure~\ref{figure:2} shows the peak region of the same resonance in greater detail, over 239--243 eV. The yield curves allow their shapes to be compared, while their ratio reveals small differences that are difficult to distinguish in an overlay. A separate SAMMY MC calculation with 10 million histories per incident energy is used for this figure.

Additional tests confirm the consistency of the calculations when a common cross-section table is used. The maximum absolute difference between the analytical $Y_0$ contribution in MC-fast and the corresponding SAMMY value is $4.8\times10^{-10}$. Increasing the number of histories reduces the scatter in the differences between the MS corrections, while repeated runs with independent random number sequences produce comparable scatter.

\subsection{Validation over 50--300 eV}

Following the local tests, the comparison is extended to 50--300 eV. At this stage, MC-fast calculates its own cross sections from the resonance parameters, while SAMMY uses its own cross-section calculation. The full calculation chain is therefore tested, from cross-section evaluation to the simulation of successive scattering events.

The independent cross-section calculation in MC-fast is checked against SAMMY over 50--300 eV. Cross sections are compared at identical energies without additional interpolation between the tables. The maximum relative difference for capture, elastic scattering and total interaction cross sections does not exceed $1.2\times10^{-4}$\%.

Over 50--300 eV, both codes calculate the capture yield on the same grid of 50 001 incident energies with a spacing of 0.005 eV. Each point uses 100 000 histories, giving approximately 5 billion histories per run. The calculations use an areal density $n_0$ = $7.3\times10^{-4}$ atoms/barn, cross sections at 295 K and a limit of five scattering events. The identical incident energy grids allow direct comparison of the calculated yields without additional interpolation.

The calculations are compared using the difference in capture yields following one or more scattering events, normalised to the total SAMMY capture yield:

\begin{equation}
r(E)=\frac{Y_{\geq1}^{\mathrm{MC-fast}}(E)-Y_{\geq1}^{\mathrm{SAMMY}}(E)}{Y_{\mathrm{tot}}^{\mathrm{SAMMY}}(E)}.
\end{equation}

\begin{figure}[H]
\centering
\includegraphics[width=1\linewidth]{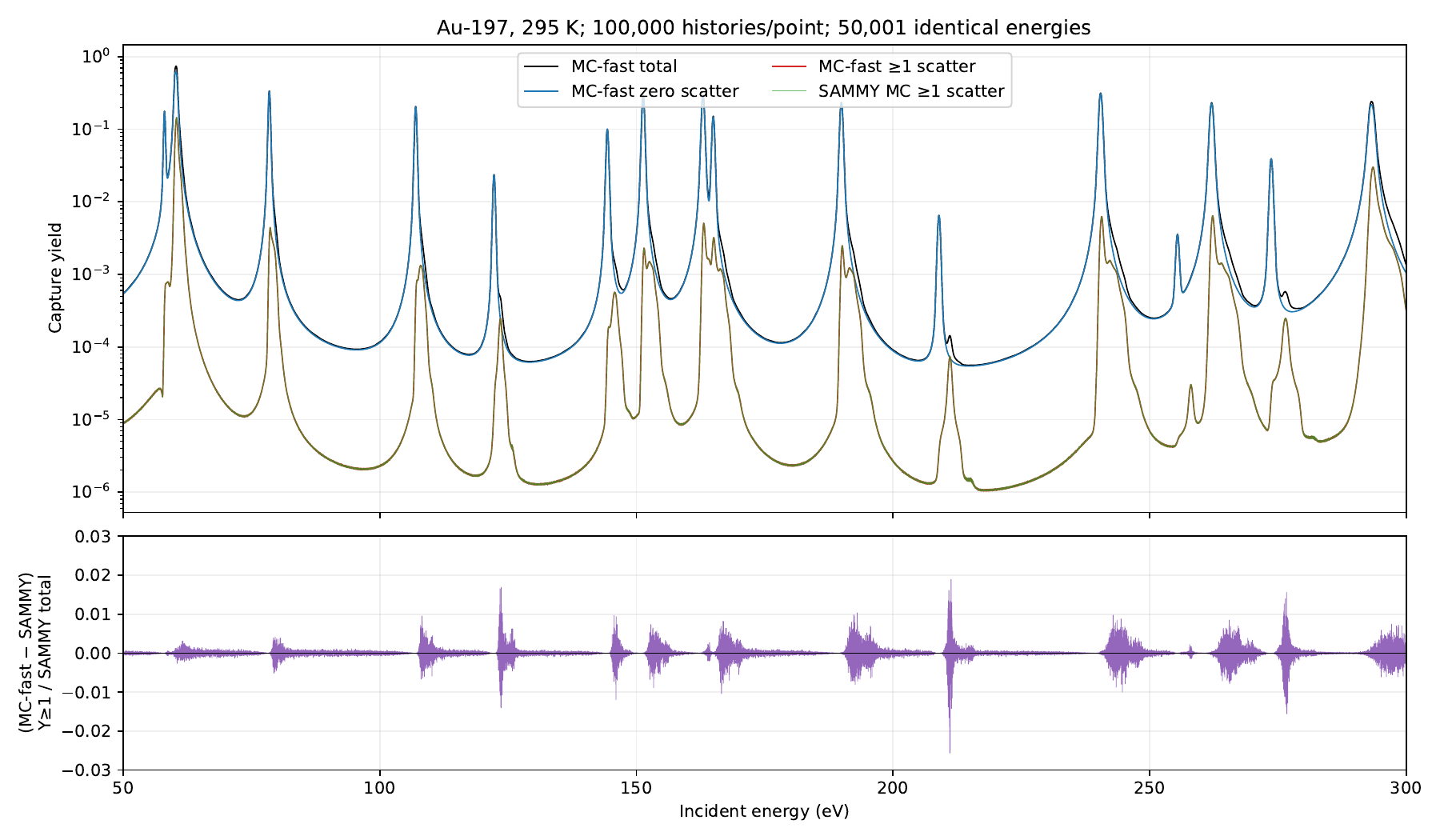}
\caption{Comparison over 50--300 eV with 100 000 histories per point. The upper panel shows $Y_{\mathrm{tot}}$ and $Y_0$ from MC-fast and $Y_{\geq1}$ from both codes. The lower panel shows the difference between the capture contributions following scattering in MC-fast and SAMMY MC, divided by the total SAMMY MC yield. The vertical range from -0.03 to +0.03 corresponds to deviations from -3\% to +3\% of the total capture yield.}
\label{figure:3}
\end{figure}

Here $Y_{\geq1}$ = $Y_1$ + $Y_{\geq2}$ and $Y_{\mathrm{tot}}$ = $Y_0$ + $Y_{\geq1}$. Thus, r(E) expresses the difference between the two calculations relative to the total capture yield. For example, r = 0.01 corresponds to a difference of 1\% of the total SAMMY yield. The upper panel of figure~\ref{figure:3} shows the capture yield curves and individual contributions, while the lower panel shows r(E), revealing local discrepancies across the full range.

Over 50--300 eV, the mean normalised difference r(E) is $6.9\times10^{-6}$ and its root mean square (RMS) is 0.00121, equivalent to approximately 0.12\% of the total SAMMY MC capture yield. This value characterises the difference across the full energy grid. At individual points, the normalised difference reaches -2.56\% and +1.89\%; therefore, 0.12\% cannot be interpreted as a bound at every point. A small mean does not exclude local deviations of opposite signs. Numerical results are given in table~\ref{table:2}.

\begin{table}[H]
\centering
\caption{Normalised difference between MC-fast and SAMMY over the broad energy range and the window near 240 eV.}
\label{table:2}
\small
\begin{tabularx}{\linewidth}{XX}
\toprule
Quantity & Result \\
\midrule
Mean r over 50--300 eV & $6.8899\times10^{-6}$ or 0.000689\% \\
RMS r over 50--300 eV & 0.0012084 or 0.12084\% \\
Minimum and maximum r & -0.025646 and +0.018946 \\
Points in the window 235.45--245.45 eV & 2001 \\
Mean r in this window & -$1.0461\times10^{-5}$ \\
Standard deviation of r in the window & 0.0017443 or 0.17443\% \\
\bottomrule
\end{tabularx}
\end{table}

\subsection{Expected statistical scatter}

In the weighted method, each history makes its own contribution to the capture yield. Precision therefore depends on both the number of histories and the spread of their contributions: for the same number of histories, uncertainties may differ at different energies. The yield is estimated as the mean contribution, and its statistical uncertainty is obtained from the variance of these contributions. Since a single history may contribute to capture after different numbers of scattering events, these contributions are correlated, and covariances must be included when estimating the uncertainty of their sum.

The statistical uncertainty is calculated from the spread of the contributions of individual neutron histories. Since the MC-fast and SAMMY MC calculations are independent, the uncertainty of their difference is

\begin{equation}
s_{\Delta}(E)=\sqrt{s_{\mathrm{MC-fast}}^2(E)+s_{\mathrm{SAMMY}}^2(E)}.
\end{equation}

where $s_{\mathrm{MC-fast}}$ and $s_{\mathrm{SAMMY}}$ are the statistical uncertainties of the calculated capture yields following one or more scattering events. The uncertainty of the normalised difference is approximately

\begin{equation}
s_r(E)\approx\frac{s_{\Delta}(E)}{Y_{\mathrm{tot}}^{\mathrm{SAMMY}}(E)}.
\end{equation}

This estimate neglects the statistical uncertainty of the denominator.

In the window 235.45--245.45 eV, the mean normalised difference between the two calculations is -$1.05\times10^{-5}$, and its standard deviation is 0.174\% of the total SAMMY capture yield. For comparison, the estimated expected statistical scatter is 0.169\%. This estimate assumes comparable statistical uncertainties in MC-fast and SAMMY MC, since SAMMY uncertainties are not provided in the output file. The proximity of these values shows that the observed scatter is consistent in magnitude with statistical noise. Quantifying a possible systematic difference requires the statistical uncertainties of both calculations.

Monte Carlo statistical noise decreases approximately as the inverse square root of the number of histories. Increasing the statistics from 100 000 to 1 million histories per point should therefore reduce the scatter in the difference between two independent calculations by approximately $\sqrt{10}$ $\approx$ 3.16. For the window near 240 eV, the expected scatter would be approximately 0.055\% of the total capture yield instead of 0.174\%. This is an estimate based on $1/\sqrt{N}$ scaling, rather than the result of a comparison between the codes at 1 million histories per point.

\subsection{Test of the scattering limit}

\begin{figure}[H]
\centering
\includegraphics[width=.88\linewidth]{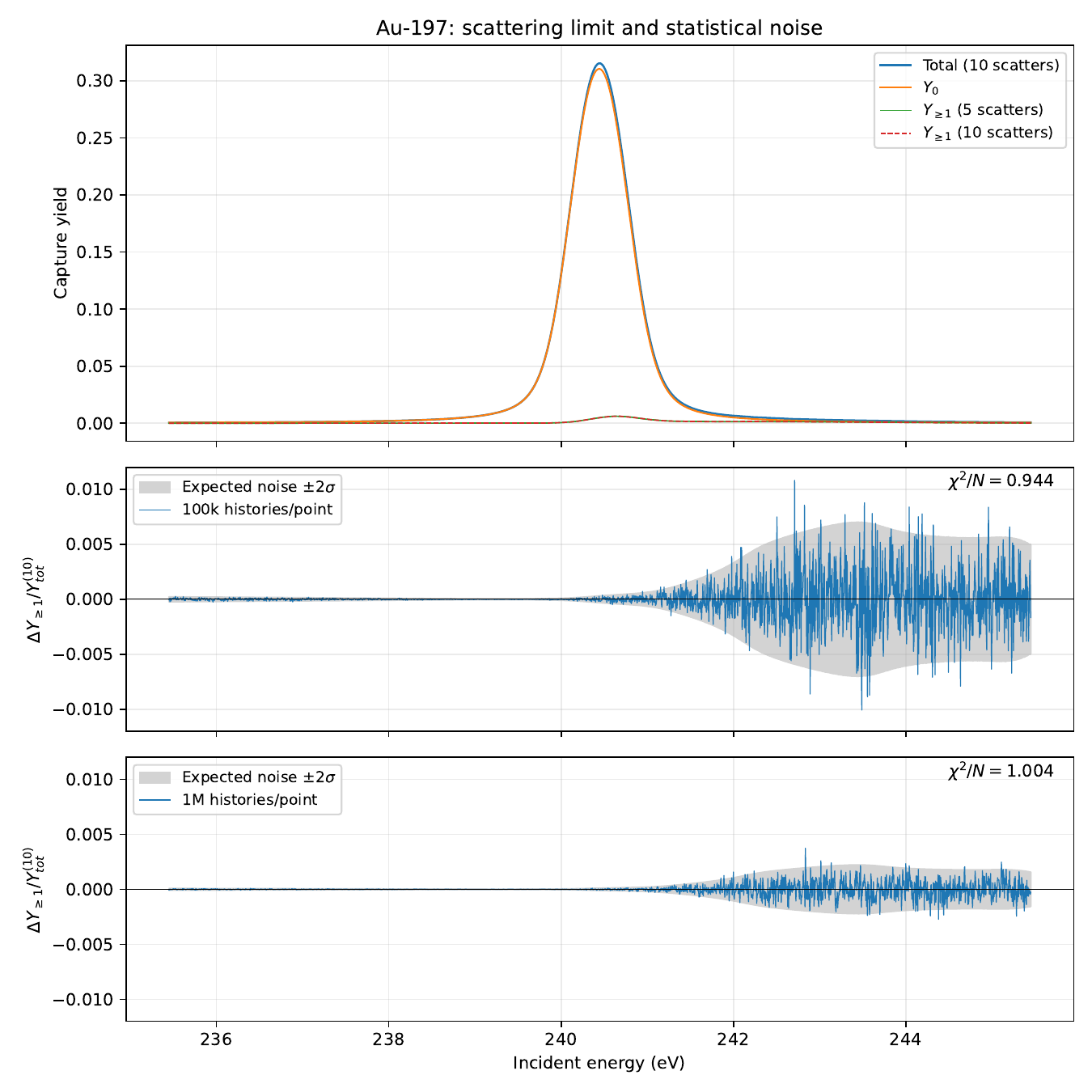}
\caption{Test of limits of five and ten scattering events near 240 eV. The upper panel shows the components at 1 million histories per point. The lower panels show the differences at 100 000 and 1 million histories on the same scale. The grey band $\pm$2$s_i$ shows the expected noise at each point. Increasing the statistics reduces fluctuations, including those above the resonance energy.}
\label{figure:4}
\end{figure}

To test the limit of five scattering events, an additional calculation with a limit of ten is performed over 235.45--245.45 eV at an areal density $n_0$ = $7.3\times10^{-4}$ atoms/barn. With 100 000 and 1 million histories per incident energy, the differences are not distinguishable from statistical noise. The difference between the results characterises the contribution from the sixth through tenth scattering events and allows the adequacy of the chosen limit to be assessed under these conditions. Contributions from higher scattering orders are not directly tested by this comparison. The results are shown in figure~\ref{figure:4}.

A higher target areal density increases the probability of repeated scattering. The comparison between limits of five and ten scattering events is therefore repeated for $n_0$ $\approx$ $6.3\times10^{-3}$ atoms/barn, corresponding to the INES example. In the same window near 240 eV, the differences are again not distinguishable from statistical noise. Increasing the number of histories from 100 000 to 1 million reduces the scatter of the difference by approximately a factor of 3.2. The results for both targets are shown in figures 4 and 5 and listed in table~\ref{table:3}.

\begin{figure}[H]
\centering
\includegraphics[width=.88\linewidth]{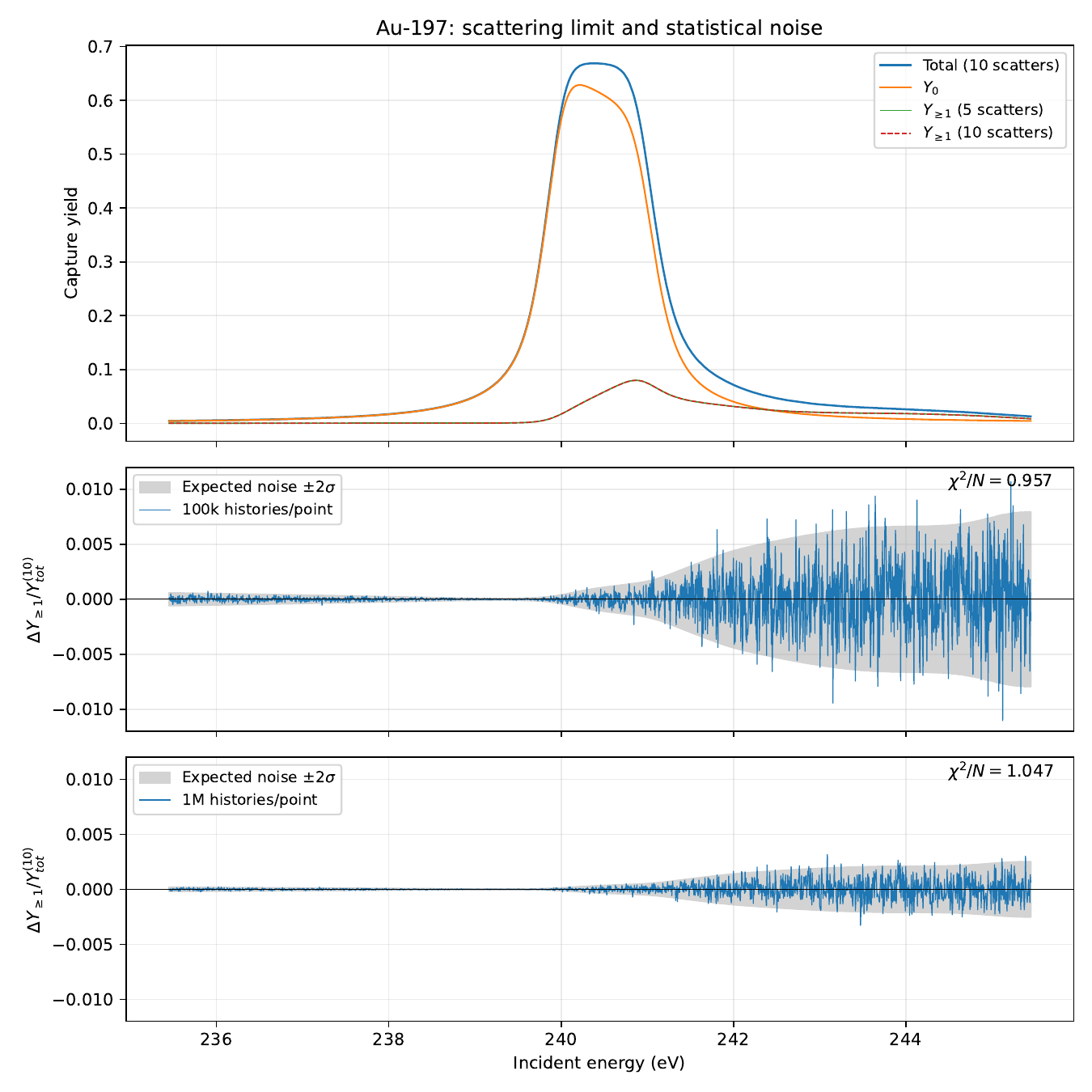}
\caption{Test of limits of five and ten at the INES areal density. The upper panel shows the capture yield contributions at 1 million histories per point. The two lower panels show the differences at 100 000 and 1 million histories on the same scale of $\pm$0.012. The grey band represents the expected noise, $\pm$2$\sigma$. Increasing the statistics reduces the scatter without revealing a persistent difference between the limits.}
\label{figure:5}
\end{figure}

The difference between the calculations is quantified using

\begin{equation}
\frac{\chi^2}{N_E}=\frac{1}{N_E}\sum_i\frac{[Y_{\geq1}^{(10)}(E_i)-Y_{\geq1}^{(5)}(E_i)]^2}{s_5^2(E_i)+s_{10}^2(E_i)}.
\end{equation}

Here $s_5$ and $s_{10}$ are the statistical uncertainties of the capture yields following one or more scattering events for the corresponding limits. The calculations with limits of five and ten use independent random number sequences, so the variance of their difference is the sum of the variances of the two calculations. The quantity $\chi^2/N_E$ is averaged over $N_E$ = 2001 energy points. A value near unity means that the mean squared difference is comparable to the sum of the variances of the two calculations. No parameters are estimated from the compared curves. In table~\ref{table:3} and figures 4 and 5, N likewise denotes the number of energy points $N_E$.

\begin{table}[H]
\centering
\caption{Test of limits of five and ten for two areal densities. RMS is calculated for $\Delta Y_{\geq1}/Y_{\mathrm{tot}}^{(10)}$ over all 2001 points.}
\label{table:3}
\small
\begin{tabularx}{\linewidth}{XXXX}
\toprule
Density and statistics & $\chi^2/N$ & RMS residual & Time for limits 5 / 10, s \\
\midrule
n\_TOF, 100 000 & 0.94398 & 0.001765 & 14.462 / 14.522 \\
n\_TOF, 1 million & 1.00372 & 0.0005392 & 126.265 / 126.258 \\
INES, 100 000 & 0.95707 & 0.001995 & 20.102 / 22.372 \\
INES, 1 million & 1.04720 & 0.0006226 & 180.245 / 203.961 \\
\bottomrule
\end{tabularx}
\end{table}

\subsection{Computing time}

Both calculations are performed on a Mac mini M4. With 100 000 histories at each of 50 001 energy points, the total wall-clock time of the MC-fast run with six worker processes is 338 s, compared with approximately 53 min for single-threaded SAMMY MC. The speedup is approximately 9.4 and includes the effect of parallel execution in MC-fast.

\section{Resonance parameter recovery and precision of the MS calculation}

Resonance parameter recovery is tested using model capture yield curves calculated with known neutron and radiative widths. During parameter recovery from these curves, the multiple scattering correction is recalculated whenever trial widths change. Comparing the recovered widths with their input values allows the effect of statistical noise in the MS calculation on parameter recovery accuracy to be assessed.

An isolated resonance near 240 eV and two nearby resonances at 163--165 eV are considered. These examples test recovery of the parameters of an individual resonance and simultaneous recovery of those of a pair. In the first case, the neutron and radiative widths of one resonance are estimated; in the second, both widths of each resonance are estimated. The input parameters are listed in table~\ref{table:4}.

\begin{table}[H]
\centering
\caption{Input energies and widths of the $^{197}$Au resonances used to calculate the reference curves. Only neutron and radiative widths are varied during recovery; resonance energies remain fixed.}
\label{table:4}
\small
\begin{tabularx}{\linewidth}{XXX}
\toprule
$E_r$, eV & $\Gamma_n$, meV & $\Gamma_\gamma$, meV \\
\midrule
162.956 & 55.160 & 129.000 \\
164.963 & 9.983 & 121.000 \\
240.447 & 74.164 & 121.000 \\
\bottomrule
\end{tabularx}
\end{table}

For the isolated resonance, an energy window of $\pm$2 eV about its energy of 240.4474 eV is used. For the pair, a window of $\pm$3 eV about the mean of the two resonance energies is used. With a spacing of 0.05 eV, these windows contain 81 and 121 points, respectively. The calculations use a target areal density $n_0$ = $7.3\times10^{-4}$ atoms/barn, a temperature of 295 K and a limit of five scattering events. Cross sections are evaluated on a grid with a spacing of 0.002 eV.

Reference curves are calculated using 1 million histories per energy point. During parameter recovery, the MS correction is calculated with fewer histories: 10 000, 30 000, 50 000 and 100 000 per point. For each number of histories, ten recovery calculations use different random number sequences in the trial MC calculation. The same reference curve is used for each example. The resulting spread of widths characterises the effect of statistical noise in the recalculated MS correction on the stability of parameter recovery. The mean deviation from the input widths is reported for that particular reference curve.

Within each recovery calculation, common random number sequences are used as the trial widths change. This reduces the random variation in differences between calculations at nearby parameter values and improves minimisation stability. The ten repeated recovery calculations use different sequences, independent of the sequence used to generate the reference curve.

Only the neutron and radiative widths of the selected resonances are varied during recovery. Resonance energies and the parameters of neighbouring resonances remain fixed, but the full resonance set is used to calculate cross sections. Initial widths are set 3\% above their input values to test recovery from a different starting point. The model curves include neither experimental background nor a resolution function.

The widths are estimated by weighted least squares. The minimised objective is the sum of squared differences between the calculated and reference capture yields, normalised by fixed uncertainty scales. At each point, the scale is the larger of the statistical uncertainty of the reference curve and $2\times10^{-4}$ times its maximum yield. This lower bound limits the weight of points with very small statistical uncertainties. Since statistical noise in the trial MS calculation is not included in the denominator, the minimised objective is not interpreted as a statistical $\chi^2$ goodness-of-fit measure for the two curves.

Minimisation uses the bounded Trust Region Reflective (TRF) algorithm implemented in scipy.optimize.least\_squares. The optimised variables are the logarithms of the ratios of trial widths to input widths; the allowed widths range from 0.7 to 1.3 times their input values.

The stopping tolerances for changes in the objective function, changes in the parameters and the gradient magnitude are set to $10^{-5}$. Derivatives are evaluated by central finite differences with a step of 0.005 in the logarithms of the width ratios. All 80 recovery calculations terminate according to the objective-function or parameter-change criterion; the recovered values remain within the specified bounds. All results are included in the estimates of mean deviations and scatter.

For each width, the relative deviation of the recovered value from the input value is calculated. The ten recovery calculations are used to obtain the mean deviation, the standard deviation of the results about their mean and the RMS relative error with respect to the input width. The standard deviation characterises the spread of repeated recovery results, while the RMS error also includes the displacement of their mean from the input value. These quantities describe the effect of MC statistical noise and are not experimental uncertainties of the widths.

For the resonance near 240 eV, increasing the number of histories reduces width recovery errors. At 10 000 histories per point, the RMS relative errors over ten recovery calculations are 0.19\% for the neutron width and 0.24\% for the radiative width. At 100 000 histories, these decrease to 0.067\% and 0.081\%, respectively. Thus, a tenfold increase in statistics reduces the errors by approximately a factor of three, close to the expected $1/\sqrt{N}$ dependence. The absolute mean relative deviations of recovered widths from their input values do not exceed 0.035\% in any of the tested series and are substantially smaller than the spread of individual results.

For the two nearby resonances at 163--165 eV, recovery errors are smaller than for the isolated resonance near 240 eV. At 10 000 histories per point, the RMS relative error of each of the four widths does not exceed 0.064\%, and at 50 000--100 000 histories it is at most approximately 0.025\%. In this example, the proximity of the resonances therefore does not prevent accurate simultaneous recovery of their widths. The difference between the isolated resonance and the pair shows that the required number of histories depends on the sensitivity of the capture yield curve to the estimated parameters.

\begin{figure}[H]
\centering
\includegraphics[width=1\linewidth]{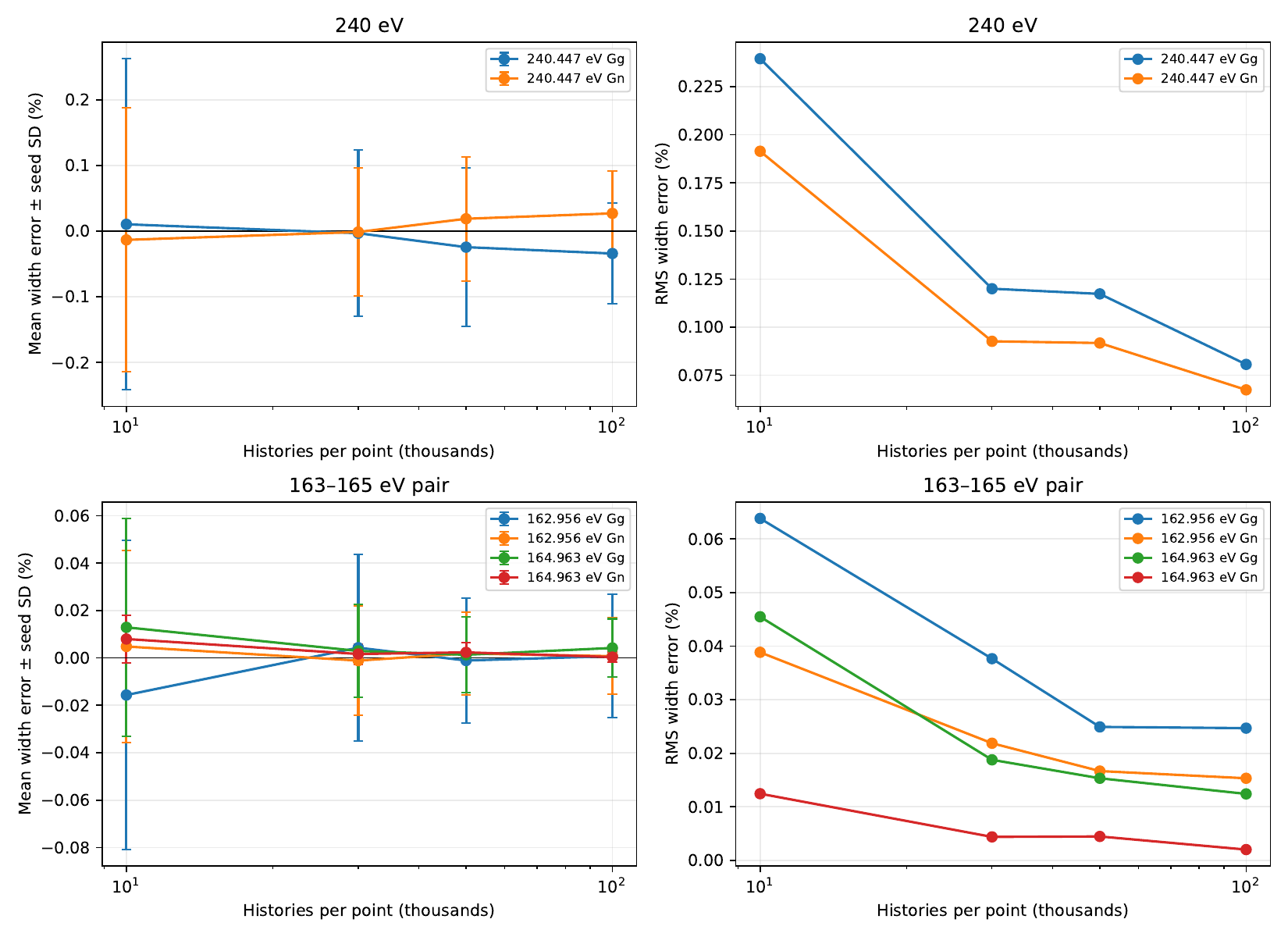}
\caption{Recovery of neutron and radiative widths for an isolated resonance near 240 eV and the 163--165 eV resonance pair at different numbers of MC-fast histories per energy point. The left panels show mean relative deviations of recovered widths from their input values; the vertical bars indicate the standard deviation of ten recovery results obtained with independent random number sequences. The right panels show RMS relative recovery errors. Each example uses one reference curve calculated with 1 million histories per point.}
\label{figure:6}
\end{figure}

On a Mac mini M4 with six worker processes, the median parameter recovery time at 30 000 histories per point is approximately 14 s for the isolated resonance and 27 s for the pair. At 100 000 histories, it increases to approximately 28 and 56 s, respectively. These times include cross-section evaluation and recalculation of the MS correction as trial widths change. Reference curve generation and plotting are excluded; the reference curve is calculated beforehand and reused in all recovery calculations. In the examples considered, width recovery with the MS correction recalculated at every change in trial parameters therefore takes tens of seconds. figure~\ref{figure:6} shows how width recovery errors depend on the number of histories.

\section{Conclusions}

The fast code MC-fast has been developed to calculate neutron capture following scattering in a target. Good agreement with an independent SAMMY Monte Carlo calculation is obtained for $^{197}$Au over 50--300 eV. In the window 235.45--245.45 eV, the observed scatter in the differences is consistent in magnitude with the estimated statistical noise. Over 50--300 eV, the RMS normalised difference between the MS contributions is approximately 0.12\% of the total SAMMY MC capture yield. Repeated calculations confirm that noise decreases as the number of simulated histories increases.

In the interval 240--241 eV, the direct MC contribution from capture following two or more scattering events is 48\%, 54\% and 72\% of the analytical SAMMY result for the three target thicknesses considered. Introducing the SAMMY approximation for neutron escape after the second scattering event into the diagnostic MC calculation gives agreement with the analytical SAMMY result, with an observed discrepancy of 0.4--1.2\%. This test links the discrepancy to the approximate treatment of neutron transport after the second scattering event.

The speed of MC-fast allows the multiple scattering correction to be recalculated whenever trial resonance parameters change. Tests with model curves for the isolated resonance near 240 eV and the 163--165 eV pair show that a few tens of thousands of histories per energy point allow widths to be recovered with small errors in tens of seconds. The tests establish how the number of MC histories affects the spread of recovered neutron and radiative widths and the computing time. These results allow the statistics used for the MS correction to be chosen according to the required parameter recovery accuracy. The estimates apply to the selected resonances, energy windows and model conditions, without experimental background or a resolution function.

\section{Acknowledgements}

The author thanks V. G. Pronyaev, I. I. Tkachev, G. K. Matushko and D. V. Hlustin for their support and assistance with this work.

\section{Statement on the use of artificial intelligence}

ChatGPT Work (OpenAI) was used as an auxiliary tool for discussing the analysis methodology, developing and checking computational programs, structuring the manuscript, translation and language editing. The author takes full responsibility for the scientific decisions, verification of calculations, interpretation of results, bibliographic references and final text, as well as the accuracy, originality and integrity of the work.

\end{document}